\documentclass[sigconf,screen]{acmart}
\usepackage{algpseudocode}
\usepackage{multirow}
\usepackage{makecell}
\usepackage{subfigure}
\usepackage{mathtools}
\usepackage{amsmath} 
\usepackage{graphicx} 
\usepackage[linesnumbered,ruled,vlined]{algorithm2e}
\usepackage{setspace}
\usepackage{listings}
\usepackage{enumitem}
\usepackage{seqsplit}

\usepackage{amssymb} 

\setenumerate[1]{itemsep=0pt,partopsep=0pt,parsep=\parskip,topsep=1pt}
\setitemize[1]{itemsep=0pt,partopsep=0pt,parsep=\parskip,topsep=1pt}
\setdescription{itemsep=0pt,partopsep=0pt,parsep=\parskip,topsep=1pt}

\usepackage[framemethod=TikZ]{mdframed}

\makeatletter
\newcommand\PROCEDURE[3][default]{%
  \ALC@it
  \algorithmicprocedure\ \textsc{#2}(#3)%
  \begin{ALC@prc}%
}
\newcommand\ENDPROCEDURE{%
  \end{ALC@prc}%
  \ifthenelse{\boolean{ALC@noend}}{}{%
    \ALC@it\algorithmicendprocedure
  }%
}
\newenvironment{ALC@prc}{\begin{ALC@g}}{\end{ALC@g}}
\makeatother

\makeatletter%
\newcommand\xlongmapsto[2][]{%
\ext@arrow 0099{\lmsfill@}{#1}{#2}}
 \def\lmsfill@{%
\arrowfill@\vdash\relbar\rightarrow}
\makeatother

\usepackage{tikz}
 
\newcommand*\Duckpost[1][1ex]{%
	\vcenter{\hbox{
        \begin{tikzpicture}
	\draw[fill] (0,0)-- (90:#1) arc (90:270:#1) -- cycle ;
	\draw (0,0) circle (#1);
	\end{tikzpicture}}
    }}
\newcommand*\Duckpre[1][1ex]{%
        \vcenter{\hbox{
	\begin{tikzpicture}
	\draw[fill] (0,0)-- (-90:#1) arc (-90:90:#1) -- cycle ;
	\draw (0,0) circle (#1);
	\end{tikzpicture}}
    }}
\newcommand*\Stop[1][1ex]{%
        \vcenter{\hbox{
	\begin{tikzpicture}
	\draw (0,0) rectangle (0.25,0.25);
	\end{tikzpicture}}
    }}
\newcommand*\Play[1][1ex]{%
        \vcenter{\hbox{
	\begin{tikzpicture}
        \draw[fill=black] (-0.1,-0.125) -- (-0.1,0.125) -- (0.1,0) -- cycle;
	\end{tikzpicture}}
    }}
\newcommand*\Pause[1][1ex]{%
        \vcenter{\hbox{
	\begin{tikzpicture}
        \draw (-0.03,-0.125) -- (-0.03,0.125);
        \draw (0.03,-0.125) -- (0.03,0.125);
	\end{tikzpicture}}
    }}
\newif\ifdraft\drafttrue
\ifdraft
\newcommand{\jinlong}[1]{\color{red} {JL: #1 :LJ} \color{black}}

\else
\newcommand{\jinlong}[1]{}
\
\fi
\AtBeginDocument{%
  }

\setcopyright{acmlicensed}
\copyrightyear{2018}
\acmYear{2018}
\acmDOI{XXXXXXX.XXXXXXX}

\acmBooktitle{Companion Proceedings of the 33rd ACM Symposium on the Foundations of Software Engineering (FSE '25), June 23--27, 2025, Trondheim, Norway}
\acmISBN{978-1-4503-XXXX-X/18/06}

\begin{document}


\title{HACMony: Automatically Detecting Hopping-related Audio-stream Conflict Issues on HarmonyOS}

\author{Jinlong He}
\affiliation{%
  \institution{Technology Center of Software Engineering, Institute of Software, Chinese Academy of Sciences}
  \city{Beijing}
  \country{China}}
\email{hejinlong@otcaix.iscas.ac.cn}
\author{Binru Huang}

\affiliation{%
  \institution{Hangzhou Institute for Advanced Study, University of Chinese Academy of Sciences}
  \city{Hangzhou}
  \country{China}}
  \email{huangbinru24@mails.ucas.ac.cn}

\author{Changwei Xia}

\affiliation{%
  \institution{Hangzhou Institute for Advanced Study, University of Chinese Academy of Sciences}
  \city{Hangzhou}
  \country{China}}
  \email{xiachangwei24@mails.ucas.ac.cn}
\author{Hengqin Yang}
\affiliation{%
  \institution{Hangzhou Institute for Advanced Study, University of Chinese Academy of Sciences}
  \city{Hangzhou}
  \country{China}}
\email{yanghq@ios.ac.cn}

\author{Jiwei Yan}
\authornote{*}
\affiliation{%
  \institution{Technology Center of Software Engineering, Institute of Software, Chinese Academy of Sciences}
  \city{Beijing}
  \country{China}}
\email{yanjiwei@otcaix.iscas.ac.cn}
\author{Jun Yan}
\affiliation{%
  \institution{Technology Center of Software Engineering, Institute of Software, Chinese Academy of Sciences}
  \city{Beijing}
  \country{China}}
\email{yanjun@ios.ac.cn}

\renewcommand{\shortauthors}{He et al.}

\begin{abstract}
HarmonyOS is emerging as a popular distributed operating system for diverse mobile devices. One of its standout features is app-hopping,
which allows users to switch apps seamlessly across different HarmonyOS devices.
However, when apps play \textit{audio-stream-hop} between different devices, they can easily trigger \textbf{H}opping-related \textbf{A}udio-stream \textbf{C}onflict (\textsf{HAC}) scenarios.
Improper resolution of \textsf{HAC} will lead to significant \textsf{HAC} issues, which are hard to detect comprehensively due to the unclear semantics of HarmonyOS's app-hopping mechanism and the lack of effective multi-app hopping testing methods.
To fill the gap, this paper introduces an automated and efficient approach to detecting \textsf{HAC} issues. 
We formalized the operational semantics of HarmonyOS's app-hopping mechanism for audio streams for the first time. Leveraging this formalization, we designed an \textbf{A}udio-stream-aware \textbf{S}tate \textbf{T}ransition \textbf{G}raph (\textsf{ASTG}) to model the behaviors of audio-streams during window transitions and proposed a model-based approach to detect \textsf{HAC} issues automatically.
Our techniques were implemented in a tool, \textsf{HACMony}, and evaluated on 20 real-world HarmonyOS apps. 
Experimental results reveal that 12 of the 20 apps exhibit \textsf{HAC} issues. Among the 53 \textsf{HAC} issues detected, a total of 18 unique \textsf{HAC} issues were manually confirmed.
Additionally, we summarized the detected issues into two typical types, namely \textsf{MoD} and \textsf{MoR}, and analyzed their characteristics to assist and guide both app and OS developers.
\end{abstract}



\keywords{HarmonyOS, Audio-Stream Conflict, App-Hopping, Mobile Testing, Large Language model}

\settopmatter{printacmref=false}
\setcopyright{none}
\renewcommand\footnotetextcopyrightpermission[1]{} 

\newcommand\Aa{{\mathcal{A} }}
\newcommand\Bb{{\mathcal{B} }}
\newcommand\Cc{{\mathcal{C} }}
\newcommand\Ee{{\mathcal{E} }}
\newcommand\Ff{{\mathcal{F} }}
\newcommand\Gg{{\mathcal{G} }}
\newcommand\Jj{{\mathcal{J}}}
\newcommand\Ll{{\mathcal{L}}}
\newcommand\Mm{{\mathcal{M} }}
\newcommand\Nn{{\mathbb{N} }}
\newcommand\Pp{{\mathcal{P} }}
\newcommand\Qq{{\mathcal{Q} }}
\newcommand\Dd{{\mathcal{D} }}
\newcommand\Ss{{\mathcal{S} }}
\newcommand\Tt{{\mathcal{T} }}
\newcommand\transet{{\mathscr{T} }}
\newcommand\Rr{{\mathcal{R} }}
\newcommand\Vv{{\mathcal{V}}}
\newcommand\Zz{{\mathcal{Z} }}

\newcommand{\HOATG}{\textsf{HOSTG}}
\newcommand\cam{{\sf cam}}
\newcommand\spk{{\sf spk}}
\newcommand\mic{{\sf mic}}
\newcommand\port{{\sf port}}
\newcommand\lan{{\sf land}}
\newcommand\event{{\sf Event}}

\newcommand{\MASS}{\textsf{MASS}}
\newcommand{\mahs}{\textsf{H}$_2$\textsf{M}}
\newcommand{\STG}{\textsf{STG}}
\newcommand{\HASTG}{\textsf{MASTG}}
\newcommand{\ASTG}{\textsf{ASTG}}
\newcommand{\tool}{\sf{HACMony}}
\newcommand{\HAC}{\textsf{HAC}}
\newcommand{\AC}{\textsf{AC}}
\newcommand{\ASC}{\textsf{ASS}}

\newcommand\HOP{{\sf StartHop}}
\newcommand\END{{\sf EndHop}}

\newcommand\flagset{\mathcal{F}}
\newcommand\bool{\mathbb{B}}
\newcommand\Int{\mathbb{Z} }

\newcommand\topapp{{\sf TopApp}}
\newcommand\getaud{{\sf GetAud}}

\newcommand\STOP{{\sf STOP}}
\newcommand\DEF{{\sf STOP}}
\newcommand\MIX{{\sf MIX}}
\newcommand\DUCK{{\sf PLAY^\downarrow}}
\newcommand\PAUSE{{\sf PAUSE}}
\newcommand\PAUSESTAR{{\sf PLAY^\Vert}}
\newcommand\PLAY{{\sf PLAY}}
\newcommand\ASS{{\sf Ass}}

\newcommand\Service{{\sf Service}}

\newcommand\rmvapp{{\textit{rmv}}}
\newcommand\addapp{{\textit{add}}}
\newcommand\getapp{{\textit{getApp}}}

\newcommand\appinst{{\sf AppInst}}
\newcommand\devinst{{\sf DeviceInst}}
\newcommand\app{{\sf App}}
\newcommand\dev{{\sf Device}}
\newcommand\prc{{\sf Proc}}
\newcommand\appstack{{\sf AppStack}}
\newcommand\hoprelation{{\sf HopRelation}}
\newcommand\devstack{{\sf DeviceStack}}
\newcommand\spdev{{\sf SupperDevice}}
\newcommand\audiostream{{\sf Audio}}
\newcommand\Window{{\sf Window}}

\newcommand\Music{{\sf MUSIC}}
\newcommand\Movie{{\sf MOVIE}}
\newcommand\Navigation{{\sf NAVIG}}
\newcommand\Communication{{\sf COMMU}}

\newcommand\MOD{{\textsf MoD}}
\newcommand\MOR{{\textsf MoR}}
\newcommand\GetASC{{\sf GetASS}}
\newcommand\ExtractElements{{\sf UnderstandWin}}
\newcommand\Verify{{\sf Verify}}
\newcommand\GetNextEvent{{\sf GetOptimalEvent}}
\newcommand\UnderstandApp{{\sf UnderstandApp}}
\newcommand\dtop{$\DUCK$$\rightarrow$$\PLAY$}
\newcommand\dtos{$\DUCK$$\rightarrow$$\STOP$}
\newcommand\stopl{$\STOP$$\rightarrow$$\PLAY$}
\newcommand\ptop{$\PLAY$$\rightarrow$$\STOP$}

\maketitle

\section{Introduction}\label{sec:intro}

The use of audio-stream is prevalent in mobile applications, covering a range of use cases from simple music playing to complex audio processing and interaction. 
When more than one apps use audio streams on a single device, their audio streams may conflict and require proper handling. 
For example, users may launch a music app to play a song first and then switch to a movie app to play a video, both of which involve audio streams.
However, if neither the music app nor the movie app handles the played audio streams according to the scenario, i.e., just let the two started audios play at the same time, users may feel confused and uncomfortable.

When there are conflicts on multiple {audio-streams}, there are specific coping solutions according to experience.
In this example, users usually expect the music-playing can be paused automatically to ensure the normal playing of the newly launched video. 
To enhance users' experience, existing mobile systems typically offer an \textit{audio-focus} feature to resolve such \textbf{A}udio-stream \textbf{C}onflicts ({\AC}s). When an app attempts to play an audio, the system requests focus for the audio stream. Only the audio stream that gains the focus can be played, i.e., if the request is rejected, the audio stream cannot be played.
If an audio stream is interrupted by another one, it loses audio focus and is expected to take actions like \textit{pause}, \textit{stop}, or \textit{lower volume} to avoid unexpected {{\AC}-related} issues.

As we can see, handling multiple audio app interaction scenarios on a single device is inherently complex. Fortunately, as apps undergo iterative updates, most app developers have made efforts to design proper and effective conflict-handling solutions for their apps.
Nowadays, with the rise of {multi-device} distributed operating systems~\cite{hw-market}, applications can not only be used on a single device but can also be migrated to other devices through hopping operations.
These emerging operating systems aim to enhance users' experience, but they also make audio app interactions scenarios much more complex.
In such a context, the existing {conflict-handling} solutions designed for single-device scenarios often lose effectiveness. 
Thus, when developing apps working on distributed operating systems, the {audio-stream} conflict handling scenarios on multiple devices should be comprehensively tested.


In recent years, as the most representative distributed mobile operating system, 
HarmonyOS has achieved remarkable success and is running on more than one billion devices~\cite{hw-data}. Developed by Huawei, HarmonyOS is a distributed platform designed for seamless integration across smartphones, tablets, smart TVs, and more.
A standout feature of HarmonyOS is \textit{app-hopping} \cite{Hopping}, a distributed operation mode that plays a fundamental role in its ecosystem. This functionality allows users to seamlessly transfer apps across different devices, enhancing convenience and flexibility. However, this innovation also complicates the resolution of {\AC}s due to the increased interplay between devices.
Through a preliminary investigation, we found that many users had complained about poor experiences caused by \textbf{Hopping-related Audio-stream Conflict} (\HAC) issues~\cite{developer1, developer2}, where the audio-stream conflicts that occur during HarmonyOS's app-hopping are improperly handled.
Given the significant disruption \HAC\ issues cause to the user experience during app-hopping across multiple HarmonyOS devices, this paper focuses on how to detect \HAC\ issues automatically and efficiently, alongside analyzing existing \HAC\ issues to provide deeper insights.

To achieve that, it is crucial first to understand how HarmonyOS's app-hopping mechanism operates and design an efficient test generation approach tailored for app-hopping scenarios.
\textbf{The first major challenge lies in the lack of semantics for the app-hopping mechanism.}
Through an investigation of the official documentation, we found that existing materials focus on highlighting the benefits of app-hopping but lack detailed descriptions of the underlying mechanism. 
This lack of clarity significantly complicates the design of effective testing approaches for app-hopping. Specifically, it increases the difficulty of determining when and how to perform hopping operations that are more likely to trigger \HAC\ issues. Moreover, this omission also impedes other hopping-related research efforts.
\textbf{The second challenge is lacking hopping-specific models designed for efficient testing.}
Although there are various models designed for mobile apps' GUI testing
~\cite{YPX13, YZW15, GSM19, MHH19, SMC17, YWY17,YLP20,LWL22, AN13,WuDYZ19,CHS18,HCW19,HWC24},
there is no {\HAC}-specific model, that describes the behaviors of audio streams of an app and can guide a compact test case generation.
Without such a model, it would be difficult to design an effective testing strategy to detect {\HAC} issues.


To address \textit{\textbf{Challenge 1}}, we meticulously picked several representative HarmonyOS native apps, designed and conducted a group of semantic experiments on app-hopping operations, and summarized the behaviors of app-hopping operations according to the experimental results.
Based on that, we first present the formalized operational semantics of HarmonyOS's app-hopping mechanism in the aspect of audio stream. 
To address \textit{\textbf{Challenge 2}}, we propose an extended FSM \cite{YPX13} called \textbf{Audio-stream-aware State Transition Graph} ({\ASTG}) to describe the behaviors of audio streams. 
Its node, \textbf{Audio-Stream-aware State} ({\ASC}), denotes the window and its audio stream status in a running app; while its edge describes the transition rule between {\ASC}s with the label of \textit{GUI events}. 
To accurately and efficiently construct {\ASTG} model, we propose an {\ASC}-targeted and LLM-driven model exploration approach, 
which adopts LLM to identify and prioritize GUI events capable of triggering audio-stream interactions, then utilizes this information 
to guide the dynamic exploration of the app under test. 
As this exploration approach can only explore the audio-stream statuses without multiple apps' interaction, we also propose an {\ASC}-guided enhancement approach to simulate the multi-app environment for extracting the extra {\ASC}s and then construct a more precise {\ASTG}.
Based on both the operational semantics of HarmonyOS's app-hopping mechanism and the fine-grained {\ASTG} model, we can finally generate targeted test cases and execute them to detect {\HAC} issues.

We implemented our proposed techniques into a tool called {\textbf{\tool}} (\textbf{H}opping-related \textbf{A}udio-stream \textbf{C}onflict issues for Har\textbf{Mony}OS) and evaluated it on 20 real-world popular HarmonyOS apps. 
The experimental results demonstrate that the proposed testing approach can efficiently detect {\HAC} issues.
Among the 20 apps, 12 were found to have {\HAC} issues. Among the 53 \textsf{HAC} issues detected, there are 18 unique {\HAC} issues, which are all manually confirmed. 
Through issue analysis, we categorized the identified {\HAC} issues into two types: \textbf{Misuse of Device} (\MOD) and \textbf{Misuse of Resolution} (\MOR). 
We further summarized their characteristics and possible causes to provide deeper insights for both application and OS developers.



The main contributions of this work are summarized as follows:
\begin{itemize}[leftmargin=1em]
\item We present the first formal semantics of the HarmonyOS app-hopping mechanism, which serves as a foundation for HAC issue testing and could inspire further research. 
\item We design the {\ASTG} model to describe the transitions of {\ASC}s in HarmonyOS apps and propose a LLM-driven dynamic exploration approach to construct {\ASTG} models. The approach is implemented into a tool {\tool}~\footnote{Available at \url{https://github.com/SQUARE-RG/hacmony}}, which is evaluated on 20 real-world apps and successfully discovered 18 unique {\HAC} issues.
\item We summarize two typical types of {\HAC} issues, namely {\MOD} and {\MOR}, and analyze their possible causes. These findings can assist and guide both app and OS developers in improving the apps' quality on distributed mobile systems.
\end{itemize}

\section{Background}\label{sec:back}

This section introduces the basic concepts around HarmonyOS apps and audio streams.
We also give a motivating example to illustrate the behavior of a real HAC issue.

\subsection{HarmonyOS: Architecture and Application}\label{sec:harmony}

\begin{figure}[!tb]
\setlength{\abovecaptionskip}{5pt}
\setlength{\belowcaptionskip}{-5pt}
    \centering
    \includegraphics[scale = 0.42]{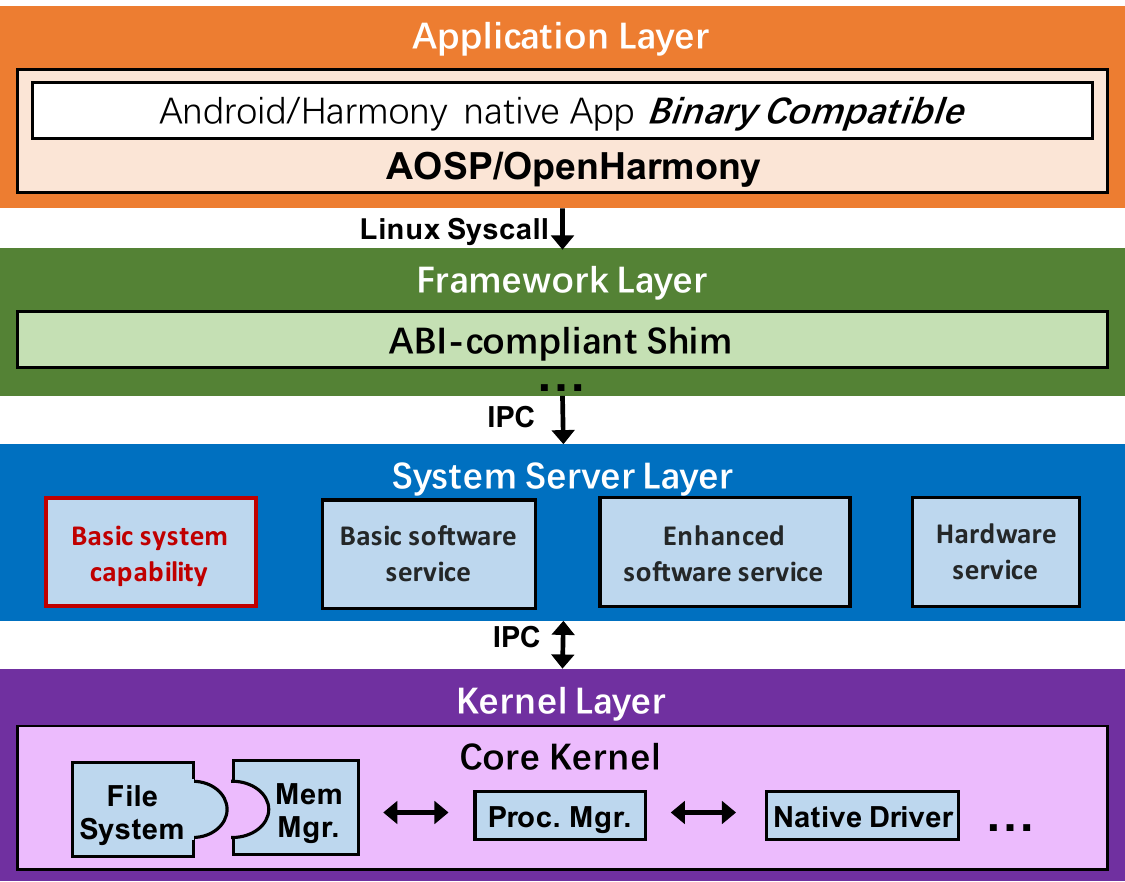}
    \caption{HarmonyOS Architecture}
    \label{fig:harmony-arch}
\end{figure}

HarmonyOS is designed with a layered architecture, which from bottom to top consists of the \textit{kernel layer}, \textit{system server layer}, \textit{framework layer}, and \textit{application layer}.
Figure~\ref{fig:harmony-arch} illustrates the layered architecture of HarmonyOS~\cite{harmonyos-arch,CMJ24}. The application layer is composed of Android (AOSP) apps and HarmonyOS native (OpenHarmony) apps, which achieves binary compatibility. 
In the framework layer, the ABI-compliant Shim (application binary interface compliant layer) redirects Linux syscalls into IPCs, channeling them towards appropriate OS services. This mechanism effectively addresses compatibility issues with AOSP and OpenHarmony, as noted in~\cite{CMJ24}.
The system service layer offers a comprehensive set of capabilities crucial for HarmonyOS to provide services to applications. It consists of a basic system capability subsystem, a basic software service subsystem, an enhanced software service subsystem, and a hardware service subsystem. 
The kernel layer, through its core kernel, furnishes memory management, file system management, process management, and native driver functionality.

With this architecture, especially the design of  \textit{ABI-compliant shim}, HarmonyOS can support both AOSP~\cite{AOSP} (for Android apps) and OpenHarmony \cite{openharmony} (for native apps).
Notably, the distributed operation \textit{app-hopping} is implemented within the \textit{basic system capability} subsystem, which transports Android and HarmonyOS native apps to another HarmonyOS device through the distributed soft bus in the same way. 
In this paper, we take both of the two supported types of apps on HarmonyOS as \textit{HarmonyOS apps}.

\subsection{Audio Stream}\label{sec:back-audio}
Audio streaming is a technology that allows users to transmit and receive audio data in real-time over the internet. Audio streaming is commonly used in online music services, internet radio, podcasts, and other applications that require instant audio content transmission. 

In general, an app typically has three \textit{fundamental} audio-stream statuses when no other app is playing audio stream, i.e., $\STOP$, $\PAUSE$, and $\PLAY$.
In real-world scenarios, audio streams are prone to conflicts when apps interact. When such conflicts occur, the behavior of the audio playback in an app becomes more complex. To mitigate the interference impact of the conflict on users, the app will often temporarily lower the volume or pause the audio stream to avoid simultaneous playback. As a consequence, during the occurrence of these conflicts, an app has two \textit{extra} audio-stream statuses, i.e., $\DUCK$ and $\PAUSESTAR$. Specifically, when an app play together with another app, $\DUCK$ signifies one app lower the volume, and $\PAUSESTAR$ signifies one app pause the playback and play again when the conflict disappears.
As shown in Table~\ref{tab:audio-status}, we consider the listed five audio stream statuses in this paper. 


\begin{table}[htbp]
\setlength{\abovecaptionskip}{5pt}
\setlength{\belowcaptionskip}{-5pt}
    \caption{Description of Audio Stream Statuses}
    \label{tab:audio-status}
    \begin{center}
    \begin{tabular}{c|c}
    \hline
        \textbf{\makecell[c]{Audio-stream Status}} & \textbf{Description}\\
    \hline
        $\STOP$ &  stop the playback\\
    \hline
        $\PAUSE$ &  pause the playback\\
    \hline
        $\PLAY$ &  play the playback\\
    \hline
        $\DUCK$ &  \makecell[c]{lower the volume, \\restore after conflict disappear} \\
    \hline
        $\PAUSESTAR$  & \makecell[c]{pause the playback, \\play after conflict disappear}\\
    \hline
    \end{tabular}
    \end{center}

\end{table}
Furthermore, HarmonyOS adopt \textit{audio focus} to manage audio streams from different apps to reconcile the audio-stream conflicts.
When an audio stream requests or releases audio focus, the system manages focus for all streams based on predefined audio focus policies. These policies determine which audio stream can operate normally and which must be interrupted or subjected to other actions. The system’s default audio focus policy primarily relies on the type of audio stream and the order in which the audio streams are initiated \cite{audio-huawei}.
In HarmonyOS, ``StreamUsage" is an enumeration type to define audio stream categories. It plays a crucial role in audio playback and management. The commonly used values include
\seqsplit{STREAM\_USAGE\_MUSIC} (MUSIC), \seqsplit{STREAM\_USAGE\_MOVIE} (MOVIE),  \seqsplit{STREAM\_USAGE\_NAVIGATION} (NAVIG), and \seqsplit{STREAM\_USAGE\_VOICE\_COMMUNICATION} (COMMU)~\cite{StreamUsage}.


Table~\ref{tab:audio-focus} lists typical resolutions for solving {\AC}s based on audio stream types by HarmonyOS, where app ``pre" plays audio streams first and then app ``post" plays at a later time.
Although these resolutions are recommended ones, HarmonyOS also allows developers to handle conflicts on their own. This leads to different proper resolutions for solving conflicts for real-world apps in practice.

\begin{table}[!hpb]
\setlength{\abovecaptionskip}{5pt}
\setlength{\belowcaptionskip}{5pt}
\caption{Typical Resolutions for Solving the {\AC}s, where 
 $\Duckpre$: app ``pre" lowers the volume, after app ``post" releases the audio focus, app ``pre" restores the volume.
 $\Duckpost$: app ``post" lowers the volume, after app ``pre" releases the audio focus, app ``post" restores the volume.
 $\Play$:  app ``pre" and ``post" play together.
 $\Pause$: app ``pre" pauses the playback, after app ``post" releases the audio focus, app ``pre" plays again.
 $\Stop$: app ``pre" stops the playback.}
\label{tab:audio-focus}
    \begin{center}
    \scalebox{1}{    
    \begin{tabular}{|c|c|c|c|c|c|}
    \hline
    \multicolumn{2}{|c|}{} & \multicolumn{4}{c|}{Type of app ``post"} \\
    \cline{3-6}
     \multicolumn{2}{|c|}{} & \Music & \Movie & \Navigation & \Communication  \\
    \cline{1-6}
    \multirow{4}*{\makecell[c]{Type \\ of \\ app \\ ``pre"}} 
    & \Music & $\Stop$ & $\Stop$ & $\Duckpre$ & $\Pause$ \\
    \cline{2-6}
    ~ & \Movie & $\Stop$ &$\Stop$ & $\Duckpre$ & $\Pause$ \\
    \cline{2-6}
    ~ & \Navigation & $\Duckpost$ & $\Duckpost$ & $\Stop$ & $\Stop$\\
    \cline{2-6}
    ~ & \Communication & $\Duckpost$ & $\Duckpost$ & $\Play$ & $\Pause$ \\
    \hline
    \end{tabular}
    }
    \end{center}

\end{table}

\subsection{Motivating Example}\label{sec:example}

\begin{figure}[!tbp]
\setlength{\abovecaptionskip}{5pt}
\setlength{\belowcaptionskip}{-5pt}
    \centering
    \includegraphics[scale = 0.25]{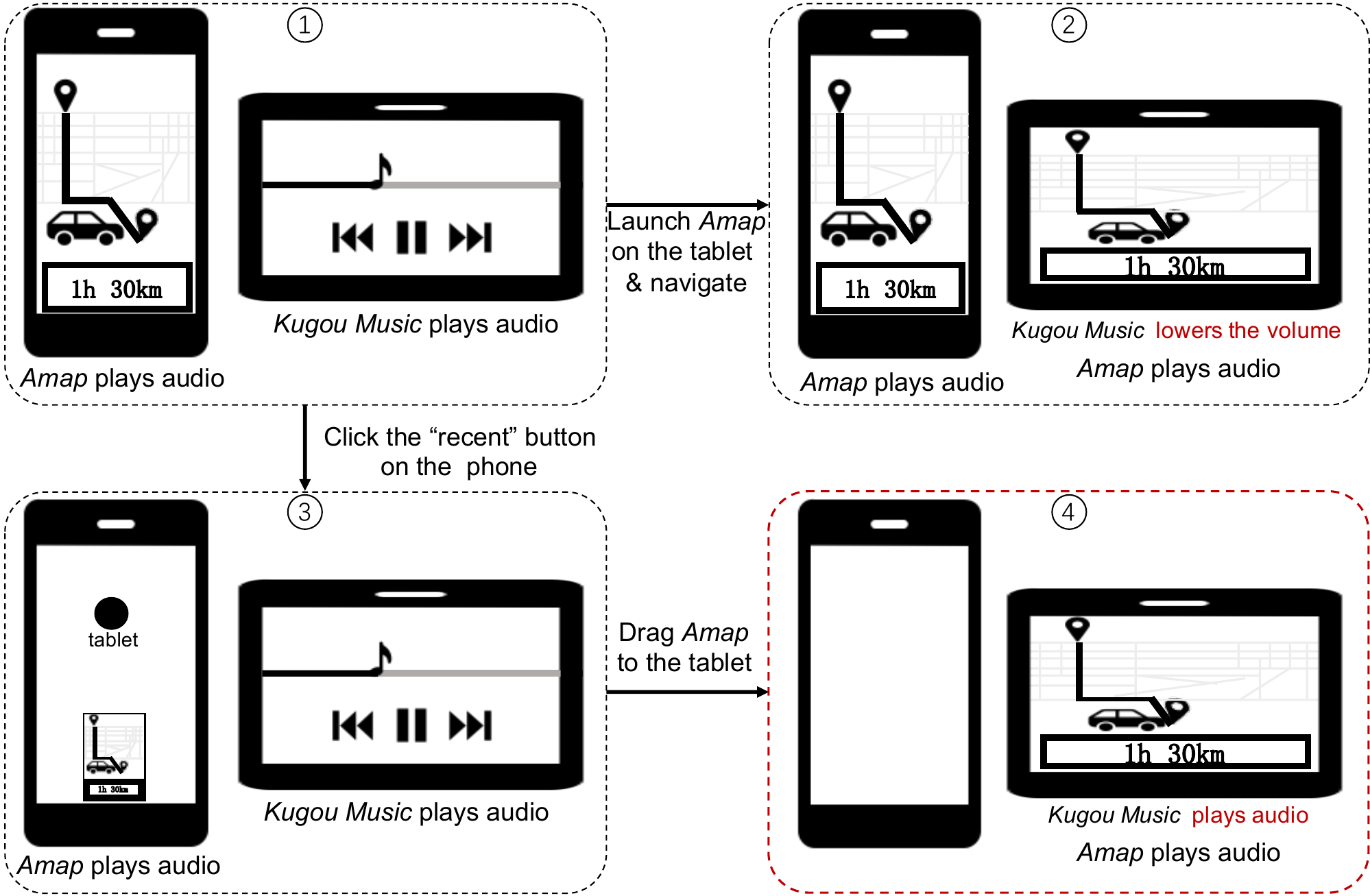}
    \caption{Motivating Example}
    \label{fig:motivating}
\end{figure}

To show the motivation for this work, we use a navigation app, \textit{AMap}~\cite{AMap}, running on a phone, and a music app, \textit{Kugou Music}~\cite{Kugou}, running on a tablet for illustration.
As shown in Figure~\ref{fig:motivating}, the initial scenario is depicted in \textcircled{1}, where both apps, \textit{AMap} and \textit{Kugou Music}, play their audio streams at normal volume.
When the user launches \textit{Amap} on the tablet and navigating in \textcircled{1}, the app \textit{AMap} plays the audio stream with the normal volume, but \textit{Kugou Music} lowers the volume to avoid the audio-stream conflict, whose status is displayed in \textcircled{2}. 
When the user clicks the ``recent" button on the phone in \textcircled{1}, the interface on the phone changes the interface for selecting the hopping app and target device, which is shown in \textcircled{3}. 
When the user drags the app \textit{Amap} to the tablet on the phone in \textcircled{3}, the app \textit{Amap} will be hopped to the tablet. 
However, in this situation, both \textit{AMap} and \textit{Kugou Music} play their audio streams at normal volume on the tablet, which is not expected. Since \textit{Kugou Music} fails to lower its volume, users may have difficulty hearing navigation instructions from \textit{AMap}. In the context of in-vehicle infotainment systems, such conflicts could even pose safety risks.

To uncover hidden vulnerabilities that can be triggered by the app-hopping operation on HarmonyOS, two key tasks must be accomplished.
First, it is essential to understand the operational semantics of HarmonyOS's app-hopping mechanism. Besides, an efficient test generation approach tailored specifically for app-hopping scenarios should be designed.





\section{The Operational Semantics of App-Hopping Mechanism on HarmonyOS  }\label{sec:hop}
In this section, we describe the overview of the app-hopping mechanism and specify the mechanism as an operational semantics.

\subsection{The Overview of App-Hopping }

HarmonyOS provides the \textit{Virtual Super Device} (Super Device) to integrate multiple physical devices and allow one device to share data and apps among devices with distributed communication capabilities. App-hopping is the fundamental feature of the Super Device to share the apps among devices~\cite{Hopping}. 
When hopping an app $a$ from device $d$ to device $d'$, the app $a$ will seamlessly transfer from device $d$ to $d'$, i.e., it will be displayed on the screen of device $d'$ only. 
Users could end a hop at any time when there is an app hop in the super device. Ending the hop of app $a$ will let app $a$ return to device $d$.
To obtain HarmonyOS's hopping mechanism, We picked several representative HarmonyOS native apps to explore the behavior of app-hooping among multiple devices.
By checking the official documents as well as conducting a group of experiments, we found that there is at most one app hop held in the super device in HarmonyOS. 
That is, if app $a$ has been hopped from device $d$ to device $d'$ and the users hop another app $a'$ in the super device, the hop of app $a$ will be ended automatically.
Furthermore, when considering the audio stream of apps, the behaviors of starting a hop and ending a hop will be more complicated.
When starting a hop of app $a$ that is playing music on device $d$ to another device $d'$, then app $a$ will play music on device $d'$. 
If there is another app playing music on device $d'$ before the hop of app $a$, the audio-stream conflict will occur on the device $d'$, which should be carefully addressed to avoid {\HAC} issue happen.

\subsection{The operational Semantics of App-Hopping }

\begin{figure}[!bp]
\setlength{\abovecaptionskip}{5pt}
\setlength{\belowcaptionskip}{5pt}
        \centering
        \includegraphics[trim = 90 0 85 0,clip,scale = 0.92]{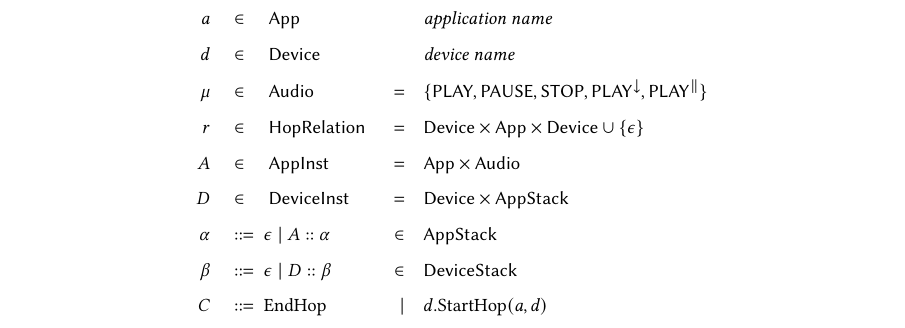}
        \caption{Domains, Stacks, and Operations}
        \label{fig:hop-syntax}
    \end{figure}

According to our literal and experimental investigation, we first summarize the formal semantics of HarmonyOS's app-hopping mechanism.
In this part, we specify its operational semantics to help users to better understand the app-hopping behaviors. 
Figure~\ref{fig:hop-syntax}~ defines domains, stacks, and operations to describe the operational semantics. We write $a$ for an app name, and $d$ for a device name.
An app instance is a pair of its activity name, 
and audio stream status $(a, \mu)$. An app stack $\alpha$ is a sequence of app instances. A device instance is a pair of its device name and its app stack $(d,\alpha)$. A device stack $\beta$ is a sequence of device instances. A hopping relation $r$ is either a triple of source device name, app name and target device name $(d,a,d')$, or a dummy symbol $\epsilon$ representing no hop exists in the super device.

The operational semantics are defined as the relation of the form $\langle \beta,r\rangle \xlongmapsto{C}  \langle\beta',r'\rangle$, where the current devices stack is $\beta$ and the current hopping relation is $r$, the operation $C$ resulting in the new devices stack $\beta'$ and the new hopping relation $r'$. 
The typical behaviors of $\HOP$ and $\END$ operations are as follows:
{
    $$\frac{
        \begin{matrix}
            \beta = \beta_1::(d_s,\alpha_s)::\beta_2::(d_t,\alpha_t)::\beta_3 \quad
            \alpha_s = \alpha_1::(a,\mu)::\alpha_2\\
            A = (a,\mu)\quad r=(d_s,a,d_t) \quad 
            \alpha_s' = \rmvapp(A,\alpha_s) \quad \alpha_t' = \addapp(A,\alpha_t)\\
         \end{matrix}}
    {\langle \beta,\epsilon\rangle \xlongmapsto{d_s.\HOP(a,d_t)} \langle\beta_1::(d_s,\alpha_s')::\beta_2::(d_t,\alpha_t')::\beta_3,r\rangle}$$

    $$\frac{
        \begin{matrix}
            \beta = \beta_1::(d_s,\alpha_s)::\beta_2::(d_t,\alpha_t)::\beta_3 \quad
            \alpha_t = \alpha_1::(a,\mu)::\alpha_2\\
            A = (a,\mu)\quad r=(d_s,a,d_t) \quad \alpha_s' = \addapp(A,\alpha_s) \quad \alpha_t' = \rmvapp(A,\alpha_t)\\
         \end{matrix}}
    {\langle \beta,r\rangle \xlongmapsto{\END} \langle\beta_1::(d_s,\alpha_s')::\beta_2::(d_t,\alpha_t')::\beta_3,\epsilon\rangle}$$
    }
The first specifies that if a user hops an app when there is no hop in the super device, the app will be moved to the target device, and the other apps in the source (resp. target) device will change the audio stream status according to the function $\rmvapp$ (resp. $\addapp$). The second describes that if a user ends a hop of an app, the behavior of this operation is dual to that of the first.

Intuitively, the function $\rmvapp(A,\alpha)$ (resp. $\addapp(A,\alpha)$) indicates the behaviors of removing (resp. adding) an app instance $A$ from (resp. into) a given app stack $\alpha$. Moreover\footnote{Due to the space limitation, we describe the remaining rules and helper functions in a companion report~\cite{semantics-pdf}.},
\begin{itemize}[leftmargin=1em]
\item if app instance $A$ is in the status $\PLAY$, and there exists another app instance $A'$ in the status $\DUCK$ or $\PAUSESTAR$, $\rmvapp(A,\alpha)$ will let $A'$ turn into $\PLAY$,
\item if $A$ is in the status of $\{\PLAY, \DUCK, \PAUSESTAR\}$, and there exists another app instance $A'$ in the status $\PLAY$, $\addapp(A,\alpha)$ will lead to the audio-stream conflict, the status of app instance $A'$ will change according to the resolution to solve the conflicts (refer to Section~\ref{sec:asc-acq}).
\end{itemize}

\begin{figure}[!t]
\setlength{\abovecaptionskip}{5pt}
\setlength{\belowcaptionskip}{5pt}
    \centering
    \includegraphics[scale = 0.37]{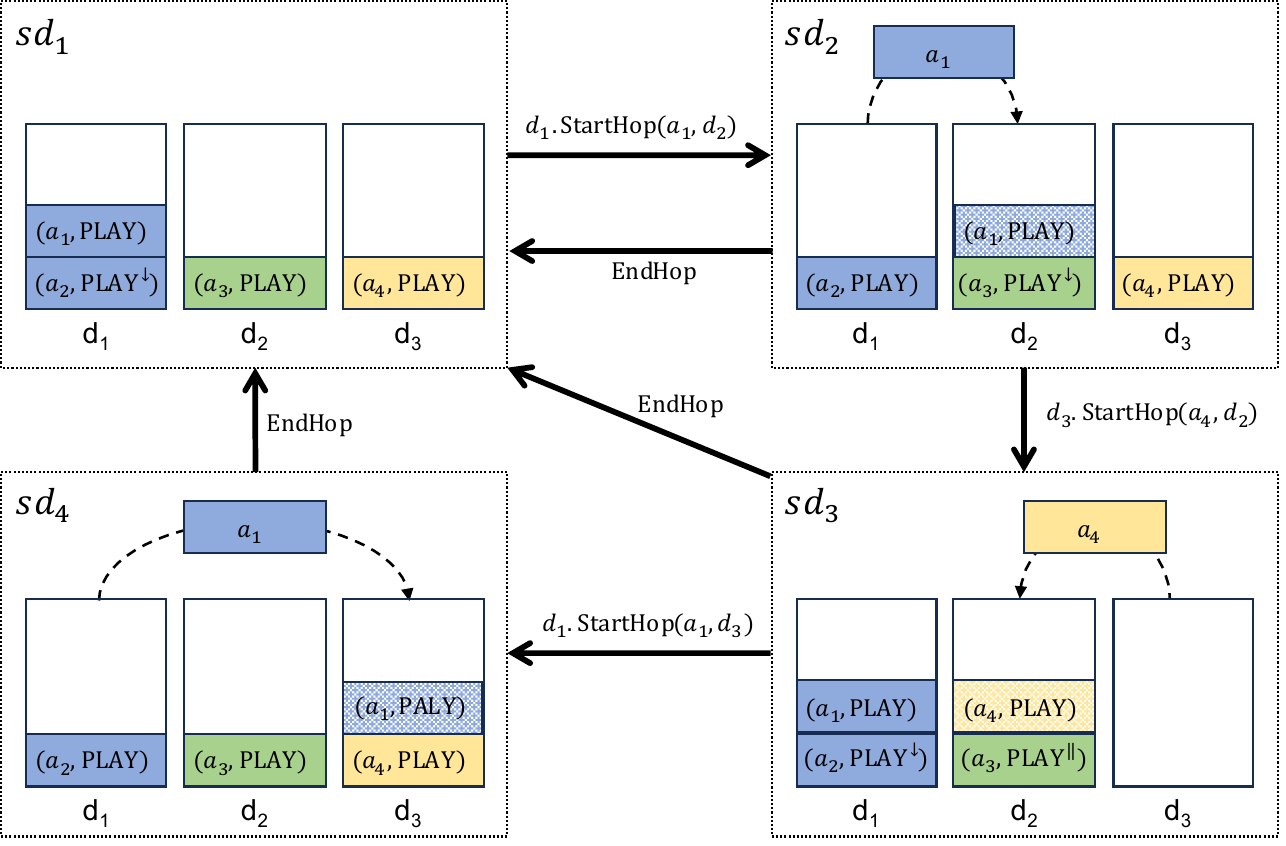}
    \caption{Example of HarmonyOS's App-Hopping Mechanism}
    \label{fig:hop-example}
\end{figure}

\textbf{Multiple-Device App-Hopping Example.} In the following, we use an example to illustrate the operational semantics of the HarmonyOS app-hopping mechanism.
Suppose that there are three devices $d_1,d_2,d_3$ in the super device, and four apps $a_1,a_2,a_3,a_4$ running on these devices. 
The types of audio streams used for each app are as follows, $a_1: \Navigation$, $a_2: \Movie$, $a_3: \Music$, and $a_4: \Communication$.
To simplify the complicated process, we suppose 
all the resolutions to solve audio stream conflicts following the typical resolutions listed in Table~\ref{tab:audio-focus}. 
As shown in Figure~\ref{fig:hop-example}, there are four cases of the super device $sd_1,sd_2,sd_3,sd_4$. For each $i\in [1,4]$, we let $sd_i = \langle \beta_i,r_i \rangle$ where 
$\beta_i = (d_1,\alpha_{i,1})::(d_2,\alpha_{i,2})::(d_3,\alpha_{i,3})$. For instance, $\alpha_{1,1} = (a_1,\PLAY)::(a_2,\DUCK)$ is the app stack of the device $d_1$ in the super device $sd_1$.
The semantics of the app-hopping mechanism are illustrated by the following cases.

\begin{itemize}[leftmargin=1em]
    \item When the operation $d_1.\HOP(a_1,d_2)$ is applied to $sd_1$, 
    the app instance $(a_1,\PLAY)$ will be removed from $\alpha_{1,1} = (a_1,\PLAY)::(a_2,\DUCK)$. Since the audio stream status of $a2$ is $\DUCK$, it will turn to $\PLAY$ 
    , resulting in $\alpha_{2,1} = (a_2,\PLAY)$. 
    Moreover, the app instance $(a_1,\PLAY)$ will be added into the device $d_2$, and request the audio focus of device $d_2$, then app instance of $a_3$ will turn to $\DUCK$, resulting in $\alpha_{2,2} = (a_1,\PLAY)::(a_3,\DUCK)$.
    \item When the operation $\END$ is applied to $sd_2$, since there is already a hop $r_2 = (d_1,a_1,d_2)$, the app instance $(a_1,\PLAY)$ will be moved back to device $d_1$ from $d_2$. Moreover,
    $(a_1,\PLAY)$ will be removed from $\alpha_{2,2} = (a_1,\PLAY)::(a_3,\DUCK)$. Since the audio stream status of $a_3$ is $\DUCK$, it will then turn to $\PLAY$, resulting in $\alpha_{1,2} = (a_3,\PLAY)$. 
    Then the app instance $(a_1,\PLAY)$ will be added into the device $d_1$, and request the audio focus of device $d_1$, then app instance of $a_2$ will turn to $\DUCK$, resulting in $\alpha_{1,1} = (a_1,\PLAY)::(a_2,\DUCK)$.
    \item When the operation $d_3.\HOP(a_4,d_2)$ is applied to $sd_2$, since there is already a hop $r_2 = (d_1,a_1,d_2)$, it will end the previous hop first. That is, the case turns to $sd_1$. Then it will hop $a_4$ from device $d_3$ to device $d_2$. Since the audio stream conflict resolution for app pair (pre:$a_3$, post:$a_1$) is different from pair (pre:$a_3$, post:$a_4$) according to their types, the audio stream status of $a_3$ is $\PAUSESTAR$ instead of $\DUCK$ in this case.
    \item When the operation $d_1.\HOP(a1,d_3)$ is applied to $sd_2$, it is similar to the previous case.
\end{itemize}

It shows that during the app-hopping, audio-stream conflicts may arise between the hopping app and audio-stream-using apps on both original device and target device, thereby altering their audio-stream statuses.


\section{Model-based Testing Approach for {\HAC} Issue Detection}\label{sec:test}

In this section, we present the overview and design details of the model-based testing approach for automatically detecting HAC issues.

\subsection{Approach Overview}\label{sec:approach-overview}
\begin{figure}[!tbp]
\setlength{\abovecaptionskip}{5pt}
\setlength{\belowcaptionskip}{5pt}
    \centering
    \includegraphics[scale = 0.34]{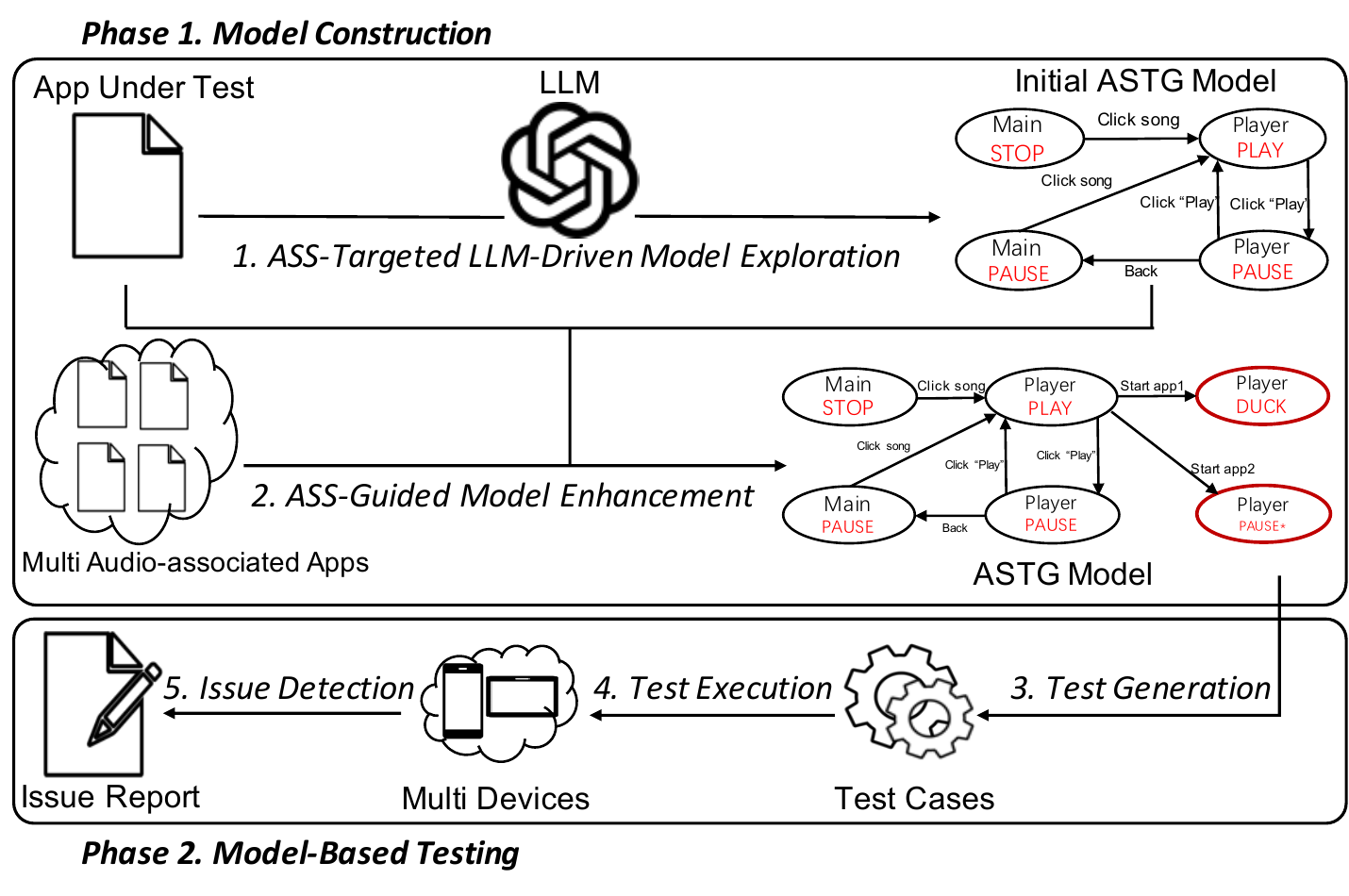}
    \caption{{\tool}'s Workflow}
    \label{fig:workflow}
\end{figure}

Based on the knowledge of the HarmonyOS's app-hopping mechanism, we design a model-based automatic testing approach for {\HAC} issue detection.
Figure~\ref{fig:workflow}~ presents an overview of {\tool}'s architecture and workflow, which has two key phases.

\textit{\textbf{Phase 1: Model Construction.}}
To obtain the GUI events that can influence the statuses of audio streams in further audio-directed testing, we designed a new model called \textbf{A}udio-stream-aware \textbf{S}tate \textbf{T}ransition \textbf{G}raph ({\ASTG}). 
The node, Audio-Stream-aware State ({\ASC}), of {\ASTG} denotes a pair of the window and its associated audio-stream status. 
This binding arises from the fact that audio stream utilization is normally achieved through windows within an app, where multiple windows may coexist to manage audio streams. Therefore, we employ {\ASC}s to explicitly define the audio-stream statuses of these windows, with the aim of testing HAC issues. 

To construct {\ASTG} model, 
We first employ a dynamic {\ASC}-targeted app exploration strategy enhanced by large language model (LLM) analysis to identify and prioritize GUI events capable of triggering audio-stream interactions. Specifically, for each single app, the LLM-driven analysis examines its window components and corresponding event handlers to identify events that may activate audio streams (e.g., play buttons, volume sliders). This refined process of identifying events is then used to construct the initial {\ASTG} model (step 1, details in Section~\ref{sec:asc-exploration}).
As the single-app exploration misses the audio-stream statuses (e.g., $\DUCK$ and $\PAUSESTAR$) that happen during the interaction of multiple apps, we enhance the initial {\ASTG} model by collaborating with multiple apps to explore extra {\ASC}s (step 2, details in Section~\ref{sec:asc-enhancement}).

\textit{\textbf{Phase 2: Model-Based Testing.}}
To generate the compact test suite for audio-aware hopping behavior testing, we select the {\ASC}s that are in the $\PLAY$-like statuses (called {\ASC}$_\PLAY$) from the {\ASTG} model constructed by Phase 1, 
and configure the devices according to different app-hopping operations (step 3, details in Section~\ref{sec:test-generation}). Then we execute the test cases on multiple devices to detect whether there are {\HAC} issues (step 4, details in Section~\ref{sec:test-execution}). 
Finally, by analyzing the resolution to solve the audio-stream conflicts during hopping and checking whether it is consistent with the resolution on a single device, {\tool} can automatically report {\HAC} issues (step 5, details in Section~\ref{sec:asc-acq}).

\subsection{{\ASTG} Model Construction}\label{sec:astg}

To generate the test case for detecting the {\HAC} issues, we define an extended FSM, 
\textbf{Audio-stream-related State Transition Graph} ({\ASTG}),
to represent the audio-stream-level behavior of an app.
An {\ASTG} model is a triple $G = (S,T,s_0)$, where
\begin{itemize}[leftmargin=1em]
    \item $S$ is a finite set of app's Audio-Stream-related States ({\ASC}s). A state $s\in S$ is a pair $\langle win,stat\rangle$ where $win$ denotes the GUI window, which contains the screenshot as well as the {\it element hierarchical tree}, $stat\in\audiostream$ denotes audio-stream status, and $s_0\in S$ is the initial {\ASC} of the app.
    \item $T$ denotes the set of transitions. An element $\tau\in T$ is a triple $\langle s, e, t\rangle$ representing the transition from the source {\ASC} $s$ to the destination {\ASC} $t$ caused by a GUI event $e$, e.g., click or drag.
\end{itemize}

\subsubsection{{\ASC}-Targeted and LLM-Driven Model Exploration}\label{sec:asc-exploration}

To conduct more effective {\ASC}-targeted GUI exploration, we utilize an LLM-driven analysis to comprehensively understand the tested app to obtain the available audio-stream types, e.g. $\Music$, and the semantics of GUI components to pick the optimal event that can enable the app to play the audio stream  corresponding to the available type.
After identifying the optimal audio-related event in current window, we proceed to execute the event on the device and collect the information about the changes (e.g., GUI window changes).
If the app has deviated from the exploration goal after previous event execution, the LLM will re-evaluate the identified event 
and select an alternative event that are more likely to lead the app towards playing the audio stream. 
This iterative process of event identification, execution, information collection, and verification continues until the app successfully plays the audio stream.


Algorithm~\ref{alg:targeted} describes the {\ASC}-targeted LLM-driven exploration approach.
The input of the approach is the {\it tested\_app} to be explored and the empty {\ASTG} $G$, the output is the {\ASTG} $G$ of the {\it tested\_app}.
First, it initializes the variable {\it feedback}, which indicates the feedback of the event execution, as empty (line 1). It also obtains the audio-stream types {\it audios} available for the {\it tested\_app} via $\UnderstandApp()$ function (line 2). 
Then, for each {\it audio} to explore from the available {\it audios}, it repeats the following process until the variable {\it feedback.terminated} becomes {\it True}, i.e., the app successfully play {\it audio} (lines 3-19).

\begin{enumerate}[leftmargin=1.5em]
    \item Obtain the current $\ASC =(win, stat)$ via the function $\GetASC()$, and 
    pass the current window $win$ to the LLM for understanding, so as to obtain a set of {\it GUI elements} containing semantic information via $\ExtractElements()$ function (lines 6-7).
    \item Send the {\it audio} to explore, the current {\it GUI elements}, the current {\ASTG} $G$, and the $feedback$ of the previous event execution, to the LLM. The LLM then selects the ``optimal'' event from all possible events based on the goal of enabling the app to play the audio stream (line 8). Then it execute the ``optimal'' event {\it event} (line 9).
    \item Obtain the current $\ASC=(win', stat')$ after executing the ``optimal'' event via $\GetASC()$ again (line 10), then update the {\ASTG} $G$ (lines 11-13).
    \item Verify whether the ``optimal'' event deviates from the exploration goal and whether the current exploration can be terminated, and record such information in the feedback. If it does deviate from the goal, then restart the {\it test\_app} to conduct a more target-directed exploration (lines 14-17).
\end{enumerate}
\begin{algorithm}
  \SetAlgoLined
  \SetKwInOut{Input}{input}\SetKwInOut{Output}{output}

  \Input{$G = (S, T, s_0), tested\_app$}
  \BlankLine

      $feedback \gets []$\;
      $audios \gets \UnderstandApp(tested\_app)$\;
      
      
    \For{each $audio$ in $audios$}{
    $feedback.terminated \gets False$\;
        \While{$feedback.terminated = False$}{
            
            $\langle win,stat\rangle \gets \GetASC()$\;
            $elements \gets \ExtractElements(win)$\;
            $event \gets \GetNextEvent($
            $audio, elements, G, feedback)$\;
            $event.\sf{execute}()$\;
            $\langle win',stat'\rangle \gets \GetASC()$\;
            $S \gets S \cup \{\langle win,stat\rangle,\langle win',stat'\rangle\}$\;
            $\tau \gets \langle \langle win,stat\rangle,event,\langle win',stat'\rangle \rangle$\;
            $T \gets T \cup \{\tau\}$\;
            $feedback \gets \Verify(stat', event,win,win')$\;
            \If{$feedback.validity$ = False}{
            Restart the $tested\_app$\;
            }
        }
    }
  \caption{Exploration()}
  \label{alg:targeted}
\end{algorithm}

\begin{table*}[!t]
\centering
\caption{The Simplified LLM Prompt Templates for ASS-Targeted Exploration}
\begin{tabular}{p{5cm}|p{5.6cm}|p{6cm}}
\hline
\textbf{$\ExtractElements$} & \textbf{$\GetNextEvent$} & \textbf{$\Verify$} \\
\hline
\textbf{Prompt:}\newline 
\textbf{Based on the information of} the provided screenshot of a mobile app interface and images of clickable components, 
\textbf{complete the task of} analyzing each component image in order to describe its function. 
\textbf{Note that,} return the answer as a Python list; each description should be concise and functional.
& \textbf{Prompt:}\newline 
\textbf{Based on the information of} the current screen, clickable elements, and previous feedback, 
\textbf{complete the task of} determining the next event.
\textbf{Note that,} focus on functionality and adapt to the current screen; choose the most appropriate element with the same purpose; avoid repeating previous events; respond only in JSON format.
& \textbf{Prompt:}\newline 
\textbf{Based on the information of} the screenshots and elements changes before and after the event,
\textbf{complete the task of} evaluating whether the event follows previous steps and progresses toward the exploration goal.
\textbf{Note that,} check if the played audio stream type matches the exploration type; assess whether to terminate based on the audio stream type and status; provide suggestions for the next step.
\\
\textbf{Example outputs:}\newline \texttt{["Return button", "Search box", "Settings button", "Add device"]} 
& \textbf{Example outputs:}\newline 
\texttt{\{"event\_type": "click", "id": 3\}\newline 
\{"event\_type": "input", "id": 2, "text": "music"\}}
& \textbf{Example outputs:} \newline 
\texttt{\{"validity": true/false,\newline
"terminated": true/false, \newline
"suggestion": "Select the "Search" button"\}} \\\hline
\end{tabular}
\end{table*}

\textbf{LLM Prompt construction}: The prompts used for interaction with the LLM in each exploration step are presented in Table 3, which will be elaborated as follows.
\begin{itemize}
    \item \textbf{$\ExtractElements$}. This prompt asks LLM to understand the semantics of each GUI element and specifies the output format. The LLM describes each element's function based on screenshots and individual element images. Each element is numbered to ensure descriptions are in order and no descriptions are missed, preventing parsing errors.
    
    \item  \textbf{$\GetNextEvent$}. This prompt asks the LLM to generate the next event based on the current app state (GUI elements and the step for exploration goal). The LLM selects the optimal event, considering previous steps and feedback to avoid repetition.
    
    \item \textbf{$\Verify$}. This prompt asks LLM to validate the executed event to checks if the event meets the exploration goal and causes GUI changes. It analyzes deviations, screen changes, and completion. It also suggests the next step to guide future event selection, and determine whether the audio-stream with the current audio type is played.
\end{itemize}

\subsubsection{{\ASC}-Guided Model Enhancement}\label{sec:asc-enhancement}

As mentioned in Section~\ref{sec:back-audio}, the audio stream statuses $\DUCK$ and $\PAUSESTAR$ occur only when there is another app requesting the audio stream focus. 
To explore the extra audio stream statuses, we need to launch another app and execute specific events to make it use the audio stream and cause audio-stream conflicts. 
For different audio stream statuses, the collaborating apps may be different in general, so we select a set of representative apps that use different types of audio streams to explore these statuses. 
The principle of the collaborating apps selection is primarily based on the typical resolutions for solving the {\AC}s (see Table~\ref{tab:audio-focus}). For example, the app with the type $\Navigation$ (resp. $\Communication$) is more likely to be selected to explore $\DUCK$ (resp. $\PAUSESTAR$) status for the app with $\Music$ type.

Algorithm~\ref{alg:extra-asc} describes the {\ASC}-guided model enhancement approach. It takes the previously constructed {\ASTG} $G = (S, T, s_0)$ by Algorithm~\ref{alg:targeted}, the previously tested app {\it tested\_app} and an audio-associated app set {\it enhanced\_apps} as inputs, and takes the {\ASTG} $G$ enhanced with extra {\ASC}s as output.
First, for the \textit{tested\_app}, it finds out all the {\ASC}s in its {\ASTG} where $stat = \PLAY$ as the target {\ASC}s. Then for each target {\ASC},  according to the {\ASTG} $G$, it obtains and executes the events to switch the \textit{tested\_app} to the {\ASC} status (lines 1-3). 
For each {\it enhanced\_app} in the audio-associated apps set {\it enhanced\_apps}, it launches {\it enhanced\_app} and switches {\it enhanced\_app} to $\PLAY$ status to make its audio stream conflict with the \textit{explored app} (lines 4-6).
Finally, if the \textit{target app} reaches a new {\ASC},
we add the new {\ASC} as well as the corresponding transition into the {\ASC} set $S$ and transition set $T$, respectively (lines 7-13).

\begin{algorithm}[!htbp]
\SetAlgoLined
  \SetKwInOut{Input}{input}\SetKwInOut{Output}{output}

  \Input{$G = (S, T, s_0), tested\_app, enhanced\_apps$ }
  \BlankLine
   
        \For{each $\langle win, stat \rangle$ in S}{
            \If{$stat = \PLAY$}{
                Switch to $\langle win, stat \rangle$\;
                \For{each $enhanced\_app$ in $enhanced\_apps$}{
                    Launch $enhanced\_app$ and switch $enhanced\_app$ to $\PLAY$ status\;
                    $\langle win',stat'\rangle \leftarrow \GetASC()$\;
                    \If{$stat \neq stat'$}{
                        $S \leftarrow S \cup \{\langle win',stat'\rangle\}$\;
                        $e \leftarrow $ launch $enhanced\_app$ and execute $enhanced\_app$\;
                        $\tau \leftarrow \langle \langle win,stat\rangle,e,\langle win',stat'\rangle \rangle$\;
                        $T \leftarrow T \cup \{\tau\}$\;
                    }
                    End $tested\_app$ and switch back to $\langle win, stat \rangle$\;
                }
                }
                }
         \caption{Enhancement()}
    \label{alg:extra-asc} 
\end{algorithm}

\subsection{Model-Based {\HAC} Issue Detection }

Upon ASTG model construction, this section presents our model-based testing approach for detecting {\HAC} issues\footnote{
We mainly consider the two-device hopping testing scenario as it is the most basic and common scenario and can cover many basic {\HAC} issues.}.  

\subsubsection{\HAC-Directed Test Generation}\label{sec:test-generation}
As we have two typical app-hopping commands $\HOP$ and $\END$, two types of hopping-related test cases $\textbf{Test}_\textbf{StartHop}$ and $\textbf{Test}_\textbf{EndHop}$ should be generated for each tested app.
The basic test generation idea is to select {\ASC}$_\PLAY$, the {\ASC}s in the $\PLAY$-like statuses in the {\ASTG} model, and configure the devices according to different app-hopping operations.
For an {\ASTG} $G = (S, T, s_0)$, an {\ASC} $= \langle win, stat \rangle \in S$ is an {\ASC}$_\PLAY$, if $stat \in \{\PLAY, \DUCK, \PAUSESTAR\}$.
Intuitively, {\ASC}$_\PLAY$ indicates the audio stream status of the window is $\PLAY$ or will turn to $\PLAY$ after other apps release the focus.

\textbf{Generate $\textbf{Test}_\textbf{StartHop}$.}
A test case $\textbf{Test}_\textbf{StartHop}$ is to perform the process of hopping the tested app where the window is in the $\PLAY$-like status to another device that is utilizing the audio stream.
With a tested app $a$ and an audio-associated app $a'$, for each {\ASC}$_\PLAY$ $s$ in the {\ASTG} of app $a$, 
we can get the following test case, $E_{s}::E_{a'}::d_1.\HOP(a, d_2)$, where 
\begin{itemize}[leftmargin=1em]
    \item $E_{s}$ is the event sequence that should be executed on device $d_1$ to let app $a$ reach {\ASC}$_\PLAY$ $s$ from the initial {\ASC} $s_0$.
    \item $E_{a'}$ is the event sequence that should be executed on device $d_2$ to let app $a'$ reach an {\ASC} whose status is $\PLAY$ from $s_0$.
    \item $d_1.\HOP(a, d_2)$ is the event that hopping the tested app $a$ from device $d_1$ to $d_2$.
\end{itemize}

\textbf{Generate $\textbf{Test}_\textbf{EndHop}$.}
A test case $\textbf{Test}_\textbf{EndHop}$ is to perform the process of ending a hop of the tested app where a window is in the $\PLAY$-like status to another device that is utilizing the audio stream.
Generating the test case $\textbf{Test}_\textbf{EndHop}$ is more complicated than $\textbf{Test}_\textbf{StartHop}$, since before ending a hop, we need to construct a hop between these two devices.
Similarly, 
a test case $\textbf{Test}_\textbf{EndHop}$ generated can be formally defined as $E_{s}::E_{a'}::\END$, where 
\begin{itemize}[leftmargin=1em]
    \item $E_{s}$ could be divided into three parts: (1) the event that starts app $a$ on the device $d_1$; (2) the $\HOP$ event that transfers app $a$ from the device $d_1$ to the device $d_2$; (3) the event sequence should be executed on device $d_2$ to let app $a$ reach {\ASC}$_\PLAY$ $s$ from the initial {\ASC} $s_0$.
    \item $E_{a'}$ is the event sequence that should be executed on device $d_1$ to let app $a'$ reach an {\ASC} which status is $\PLAY$ from $s_0$.
     \item $\END$ is the event that ending the hop of $a$ between $d_1$ and $d_2$.
\end{itemize}

\subsubsection{\HAC-Directed Test Execution}\label{sec:test-execution}

After test generation, {\tool} connects two devices  $d_1$ and $d_2$ via HarmonyOS Device Connector~\cite{hdc} (HDC) or Android Debug Bridge~\cite{adb} (ADB) to automatically execute the test cases for the target HarmonyOS app. 
For general click events or the $\END$ operation, {\tool} directly invokes the click command in HDC (or ADB) to execute the event. 
Note that, the $\END$ operation can also be regarded as a click event. 
The $\HOP$ operation could be regarded as a sequence of events, {\tool} needs to click the ``Recent" button, and then drag the current app to the target device. Finally, {\tool} records the {\ASC}s of the tested app $a$ and the conflicting app $a'$.

\subsubsection{{\HAC} Issue Detection}\label{sec:asc-acq}
After each test execution, {\tool} will analyze the recorded {\ASC}s, and report how the hopping-related audio-stream conflict between apps is resolved, i.e., the conflict resolution. 
To detect the {\HAC} issues, 
our key idea is that the conflict resolutions that show up in the multiple-device scenario, i.e., the ``hopping" resolutions, should be consistent with the ones in the single-device scenario, i.e., the ``normal" resolutions.
Thus, for each target app $a$ and its collaborating app $a'$ in app-hopping testing, for each {\ASC} $s = \langle win, \PLAY\rangle$ (resp. $s' = \langle win', \PLAY\rangle$) in the {\ASTG}, we perform the following operations to obtain the ``normal" resolutions:
\begin{enumerate}[leftmargin=1.5em]
    \item Start app $a$ on the device, and execute it to the {\ASC} $s$;
    \item Start app $a'$ on the device, and execute it to the {\ASC} $s'$;
    \item Obtain the current {\ASC}s for app $a$ and $a'$.
\end{enumerate}
We compare the ``normal" resolutions with the ``hopping" resolutions obtained by {\tool}'s test execution. If there is any inconsistency, a {\HAC} issue will be reported.

\section{Evaluation}\label{sec:exp}
To evaluate the effectiveness of our approach, we raise several research questions as follows:
\begin{itemize}[leftmargin=1em]
\item \textbf{RQ1 ({{\ASTG} Construction})} Is the {\ASTG} construction process  effective and efficient?
\item \textbf{RQ2 ({\HAC} Issue Detection)} To what extent can {\tool} detect real-world {\HAC} issues?
\item \textbf{RQ3 ({\HAC} Issue Analysis)} What are the categories and characteristics of the detected {\HAC} issues?
\end{itemize}

\subsection{Evaluation Setup}\label{sec:setup}
To answer these research questions, we collect 20 real-world HarmonyOS apps from Huawei AppGallery~\cite{huawei-app}. More specifically, we take four categories associated with audio, i.e., Music, Video, Navigation, and Social, into account, which are respectively have the highest possibility of using the audio stream type $\Music$, $\Movie$, $\Navigation$, and $\Communication$.
For each type, we download its top five apps that support app-hopping as well as available for both phone and tablet versions.
Table~\ref{tab:apps}~lists the detailed information of these experimental apps.
All of the following experiments are done on a phone HUAWEI P40 Pro and a tablet HUAWEI Matepad, both with HarmonyOS 4.2.

\begin{table*}[htbp]
    \setlength{\abovecaptionskip}{5pt}
    \setlength{\belowcaptionskip}{5pt}
        \caption{Experimental Apps and Model Size }
    \begin{center}
    \begin{tabular}{c|c|c|c|c|c|c|c|c|c}
    \hline
        \textbf{App name } & \textbf{Categories}& \textbf{Size(MB)} & 
        \textbf{Version} & 
        \makecell[c]{\textbf{\#Audio-Type}} &
        \makecell[c]{\textbf{\#ASS-}\\\textbf{Init}} &
        \makecell[c]{\textbf{\#ASS-}\\\textbf{Extra}} &
        \textbf{\#ASS}&
        \textbf{\#Edge} & \textbf{Time(s)}\\
    \hline
    Kugou Music	&Music	&156.6	&
    12.4.2 &
    2&
    4 & 
    4& 
    8& 
    7& 
    95\\
    QQ Music	&Music	&188.7	&
    13.9.0.8 &
    2&
    5 & 
    4& 
    9 & 
    8& 
    121\\
    Kuwo Music	&Music	&181.4	&
    11.0.0.0 &
    2&
    4 & 
    4& 
    8& 
    7& 
    85\\
    Fanqie	&Music	&71.5	&
    7.41.18 &
    1&
    2 & 
    2& 
    4& 
    3& 
    27\\
    Kuaiyin	&Music	&75.8	&
    5.57.11 &
    1&
    2 & 
    2& 
    4& 
    3& 
    25\\
    \hline
    Tencent Video&Video&145.9   &
    8.11.71 &
    1&
    2 & 
    2& 
    4& 
    3&
    40\\
    Xigua Video&Video&65.6  &
    8.8.6 &
    1&
    1 & 
    2& 
    3& 
    2&
    22\\
    Youku Video&Video&123.5 &
    11.0.99 &
     1&
    2 &
    2& 4& 3&38\\
    Mangguo TV&Video&133.2  &
    8.13.0 &
    1&
    2 &
    2& 4& 3&41\\
    HaoKan Video&Video&49.5 &
    7.64.0.10 &
    1&
    2 & 
    2& 
    4& 
    3&22\\
    \hline
    AMap&Navigation&254.9   &
    15.01.0 &
     1&
    4 & 
    1& 
    5& 
    4&135\\
    Baidu Map&Navigation&171.5  &
    20.7.30 &
    1&
    4 & 
    1& 
    5& 
    4&59\\
    Tencent Map&Navigation&162  &
    10.11.1 &
    1&
    4 & 
    1& 
    5& 
    4&62\\
    Petal Maps&Navigation&83.9  &
    4.5.0.303 &
     1&
    5 & 
    1&
    6& 
    5&
    74\\
    Beidouniu&Navigation&59.1   &
    3.3.1 &
    1&
    4 & 
    1& 
    5& 
    4&
    61\\
    \hline
    Douyin&Social&271.9 &
    31.4.0 &
     2&
    6 & 
    3& 
    9& 
    8&
    106\\
    Soul&Social&158.5   &
    5.40.0 &
    2&
    7 & 
    3& 
    10& 
    9&
    120\\
    Xiaohongshu&Social&164  &
    8.52.0 &
    2&
    7 & 
    3& 
    10& 
    9&
    192\\
    Zhihu&Social&87.8   &
    10.22.0 &
    1&
    3 & 
    2& 
    5& 
    4&53\\
    Momo&Social&127 &
    9.13.10 &
    2&
    6 & 
    3& 
    9& 
    8&107\\
    \hline
    \textbf{Avg./Max.}&-&\textbf{136.6/271.9}&-&
    \textbf{1.4/2}&
    \textbf{3.8/7}&\textbf{2.3/4}&\textbf{6.1/10}&\textbf{5.1/9}&\textbf{72/192}\\\hline
    \end{tabular}
    \end{center}
    \label{tab:apps}
\end{table*}


\subsection{RQ1: {{\ASTG} Construction}}\label{sec:rq1}
Table~\ref{tab:apps} shows the results of the {\ASTG} model of each experimental app constructed by {\tool}.
The fifth column gives the number of audio-stream types that are analyzed by LLM (\textbf{\#Audio-Type}).
The sixth and seventh columns give the statistics of the {\ASC}:
the number of {\ASC}s that are detected by exploration (\textbf{\#{\ASC}-Init}), and the number of the extra {\ASC}s (\textbf{\#{\ASC}-Extra}) extracted by collaborating with multiple apps.
Besides, the number of total {\ASC}s, the edges in the model, and the dynamical exploration time are shown in the last three columns.

As we can see, {\tool} 
can successfully explore all apps with an average of 1.4 audio-stream types and 6.1 {\ASC}s in an average of 72 seconds.
Additionally, 7 (35\%) apps with Music or Social category have discovered 2 audio-stream types, and the types of these apps additionally found are all $\Movie$ type.
Among all categories, the Navigation apps require more exploration time, which typically explore around 4 {\ASC}s (GUI windows).
This is because Navigation app usually involves complex events such as selecting a destination and means of transportation before starting navigation. This observation also demonstrates the effectiveness of the LLM-based exploration approach. 
Furthermore, the number of the extra {\ASC}s extracted by the {\ASC}-guided enhancement is twice the number of the audio-stream types in all apps with the Music or Video categories, indicating that these apps all discovered the $\DUCK$ and $\PAUSESTAR$ statuses during the enhancement process.

\subsection{RQ2: {\HAC} Issue Detection}\label{sec:rq2}

Table~\ref{tab:detection} displays information of the real-world {\HAC} issues detected by {\tool}. 
Columns \textbf{\#Test Cases} and \textbf{Avg. L} show the number of test cases and their average length. Columns \textbf{\#\HAC} and \textbf{\#Unq.~\HAC} show the number of the total and unique  {\HAC} issues detected. And the column \textbf{Time} shows the time of testing.

In total, with the {\ASTG} model, {\tool} generates an average of 137 test cases for each app, with an average length of 6.1 events. 
There are 12 out of 20 (60\%) apps detected to have {\HAC} issues, which involve a total of 18 unique {\HAC} issues out of 53 {\HAC} issues. This indicates that {\HAC} issues are relatively likely to occur during the HarmonyOS app-hopping. The video demonstrations of {\HAC} issues found by {\tool} can be viewed ~\cite{Youtube}. 

Recall that during the exploration phase, we leverage large models to explore multiple audio-stream types (MT) within apps, while in the enhancement phase, we utilize other apps to explore multiple audio-stream statues (MS), e.g., $\DUCK$ and $\PAUSESTAR$, with the aim of discovering as many $\PLAY$-like {\ASC}s as possible to detect more {\HAC} issues.
To validate the effectiveness of MT and MS in our approach, we conducted additional experiments, as shown in Table~\ref{tab:insertion}. The second to fifth columns shows the number of {\HAC} issues and unique {\HAC} issues of each configuration, specifically,
\begin{itemize}[leftmargin=1em]
    \item The \textbf{Base} column represents the baseline configuration without using either MS or MT.
    \item The \textbf{MT} column denotes the configuration using only multi audio-type (without MS).
    \item The \textbf{MS} column denotes the configuration using only multi audio-status (without MT).
    \item The \textbf{MT+MS} column represents the full configuration  that combines both MS and MT.
\end{itemize}
The results demonstrate that the full configuration (MT+MS) outperforms all other configurations, with a 35.9\% increase in detection of the {\HAC} issues and a 12.5\% increase in detection of the unique {\HAC} issues compared to the baseline.
Individually, MT and MS improve the detection of the {\HAC} issues by 25.6\% and 7.7\% over the baseline, respectively, but show limited improvement in the detection of the unique {\HAC} issues (only one new unique {\HAC} issue each). 
This reflects that both the MT- and MS-aware exploration are important in achieving comprehensive {\HAC} issues detecting. Testers should focus their efforts on generating test cases for different audio-types and audio-statuses to detect {\HAC} issues.


\begin{table}[!h]
    \setlength{\abovecaptionskip}{0pt}
    \setlength{\belowcaptionskip}{5pt}
    \caption{Detected {\HAC} Issues by {\tool}}
    \begin{center}
    \scalebox{0.9}{ 
    \begin{tabular}{c|c|c|c|c|c}
    \hline
        \textbf{App name} & \makecell[c]{\textbf{\#Test} \\ \textbf{Cases}}& \textbf{Avg. L}& \makecell[c]{\textbf{\#HAC}} & \makecell[c]{\textbf{\#Unq.}\\\textbf{HAC} } & \textbf{Time(s)} \\
    \hline
    Kugou Music	&228	& 5.7& 0& 0 & 2891\\
    QQ Music	&228	& 6.7&	3& 1& 3402\\
    Kuwo Music	&228	& 6.2& 5& 1 & 3202\\
    Fanqie	&114	& 5.2&	0&  0 & 1407\\
    Kuaiyin	    &114	& 6.2& 0& 0 & 3221\\
    \hline
    Tencent Video   &114& 5.2& 5& 2& 1337\\
    Xigua Video     &114& 6.2& 0& 0 & 1634\\
    Youku Video     &114& 5.2& 5 & 1 & 1367\\
    Mangguo TV        &114& 5.4& 0& 0 & 1417\\
    HaoKan Video    &114& 5.2& 4 & 1 & 2793\\
    \hline
    AMap        &76  & 7.3& 7&  3& 1241\\
    Baidu Map   &76  & 7.3& 5& 2 & 1375\\
    Tencent Map &76  & 8.4& 2& 2 & 1187\\
    Petal Maps  &76 & 7.3& 5& 2 & 1793\\
    Beidouniu   &76 & 7.1& 3 & 1 & 1857\\
    \hline
    Douyin      &190& 5.6& 0& 0 & 1450\\
    Soul        &190& 5.4& 0& 0 & 1415\\
    Xiaohongshu &190& 5.4& 3& 1 & 1453\\
    Zhihu       &114& 5.4& 0& 0 & 1417\\
    Momo        &190& 6.2& 5 & 1 & 1572\\
    \hline
    \textbf{Avg.}     &\textbf{137}& \textbf{6.1}&\textbf{2.7 }&\textbf{0.9}&\textbf{1872}\\
    \hline
    \end{tabular}
    }
    \end{center}
    
    \label{tab:detection}
\end{table}

\begin{table}[!htbp]
    \setlength{\abovecaptionskip}{0pt}
    \setlength{\belowcaptionskip}{5pt}
    \caption{Impact of Multi Audio-Type (MT) and Multi Audio-Status (MS) on {\HAC} Detection}
    \begin{center}
    \scalebox{1}{ 
    \begin{tabular}{c|c|c|c|c}
    \hline
        \multirow{2}{*}{\textbf{App name}} & \multicolumn{4}{c}{\textbf{\#{\HAC} (Unq.)}}\\
    \cline{2-5}
        & \textbf{Base}  & \makecell[c]{\textbf{MT}} & \makecell[c]{\textbf{MS}} & \makecell[c]{\textbf{MT+MS}}\\
    \hline
    QQ Music	& 2 (1)& 2 (1) & \textbf{3 (1)} &3 (1)\\
    Kuwo Music	& 5 (1) &5 (1) & 5 (1) & 5 (1)\\
    \hline
    Tencent Video   & 3 (1) & 3 (1) & \textbf{5 (2)} & 5 (2)\\
    Youku Video   & 5 (1) & 5 (1) & 5 (1) & 5 (1)\\
    HaoKan Video   & 4 (1) & 4 (1) & 4 (1) & 4 (1)\\
    \hline
    AMap        & 4 (3) & \textbf{7 (3)} & 4 (3) & 7 (3)\\
    Baidu Map   & 3 (2) & \textbf{5 (2)} & 3 (2) & 5 (2)\\
    Tencent Map & 2 (2) & 2 (2) & 2 (2) & 2 (2)\\
    Petal Maps  & 3 (2) & \textbf{5 (2)} & 3 (2) & 5 (2)\\
    Beidouniu  & 3 (1) & 3 (1) & 3 (1) & 3 (1)\\
    \hline
    Xiaohongshu  & 0 (0) & \textbf{3 (1)} & 0 (0) & 3 (1)\\
    Momo  & 5 (1) & 5 (1) & 5 (1) & 5 (1)\\
    \hline
    \textbf{Sum.}&\textbf{39 (16)}&\textbf{49 (17)} & \textbf{42 (17)} & \textbf{53 (18)}\\\hline
    \end{tabular}
    }
    \end{center}
    \label{tab:insertion}
\end{table}

\subsection{RQ3: {\HAC} Issue Analysis}\label{sec:rq3}
To assist both the developer of Harmony apps and OS better understanding the real-world {\HAC} issues. We category issues and perform case studies to investigate their characteristics.

First, we summarize the specific behaviors of the apps where {\HAC} issues occur and category issues into two types, \textbf{Misuse of Device ({\MOD})} and \textbf{Misuse of Resolution ({\MOR})}. 
{\MOD} issue refers to the situation where, during the hopping of an app, the usage of the audio streams fails to be transferred to the target device along with the app. The {\MOR} issue refers to the situation where, during the hopping of an app, an audio-stream conflict occurs on the target device, but the ``normal" resolution to solve the conflict is not applied.
In our experiments, {\tool} detected four apps with {\MOD} issues and nine apps with {\MOR} issues.

\begin{figure}[htpb]
        \centering
        \subfigure[Issue-related apps and unique issues]{   
            \label{fig:hac}
        \centering    
        \includegraphics[scale=0.4]{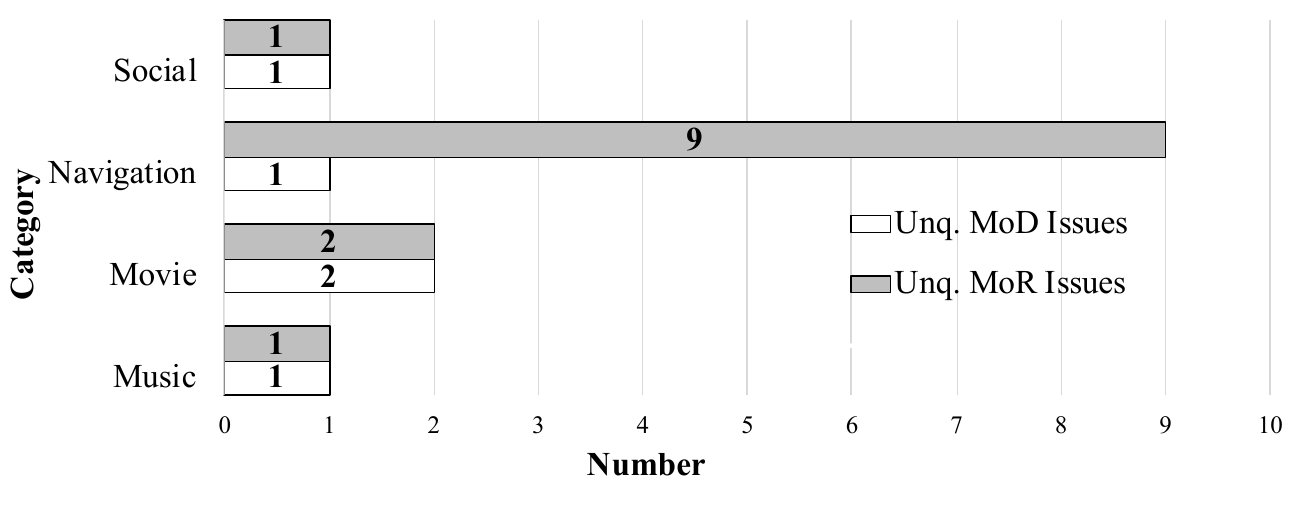}  
        }
        \subfigure[Issues triggered with different ops]{   
            \label{fig:opt}
        \centering    
        \includegraphics[scale=0.4]{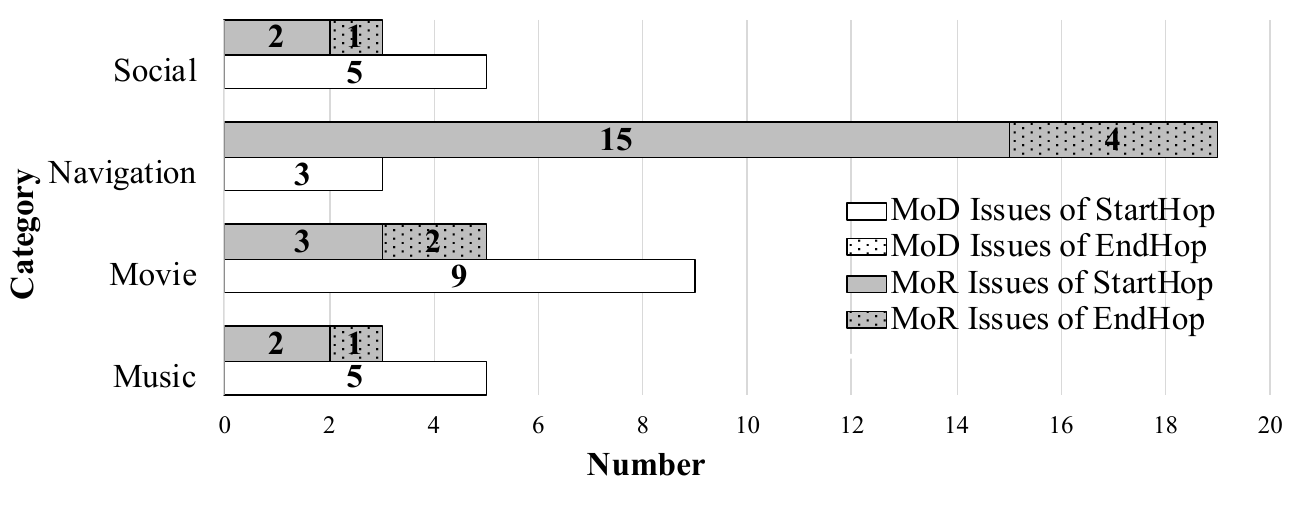}  
        }
        \caption{Number of {\HAC} Issues}
        \label{fig:hac-number}
\end{figure}

Then, we count the number of {\HAC} issues of different app categories.
As shown in Figure~\ref{fig:hac}, the {\MOD} issues are more likely to occur in the Video applications, while the {\MOR} issues are more likely to occur in the Navigation applications.
Furthermore, we count the number of {\HAC} issues of different types of test cases. As shown in Figure~\ref{fig:opt}, 
all the {\MOD} issues are detected through the test cases in the form of $\textbf{Test}_\textbf{StartHop}$, and few (26\%) {\MOR} issues are detected through the test cases in the form of $\textbf{Test}_\textbf{EndHop}$.
Although most of the {\HAC} issues are detected through the $\textbf{Test}_\textbf{StartHop}$ test cases, there are still some issues identified by the $\textbf{Test}_\textbf{EndHop}$ test cases, this indicates that it is necessary to consider different operations when generating test cases (See Section~\ref{sec:test-generation}).
Therefore, Navigation apps trigger more {\HAC} issues than all other types. They suffer severe {\MOR} issues, especially in the process of \textsf{StartHop} operation. Besides, Video apps are easier to trigger {\MOD} issues. \textit{Testers} and \textit{developers} can perform testing/developing according to the type of the target app.


To further study the characteristics of {\HAC} issues, we analyze the two types of {\HAC} issues with case studies, respectively.
\subsubsection{Misuse of Device: {\MOD} issues}
According to Finding 2, we pick a Video app to investigate the characteristic of {\MOD} issue.

\textbf{Case study 1:} {\MOD}. When \textit{Youku Video}~\cite{youku} is playing a video normally on the mobile phone, if the user hop it to the tablet, the video continues to play on the tablet, but the audio is still playing on the mobile phone. It leads to the audio-visual inconsistency problem which makes it difficult for the users to focus on the video content and affects users' understanding and enjoyment of the video.

\textbf{Analysis: } We noticed that, the {\MOD} issues may \textbf{not} occur in all test cases of the same app, i.e., sometimes the issue do not occur. Thus, we infer that such sporadic issues may be caused by the lack of synchronization of commands and data between devices, which prevents the new device from taking over audio playback in a timely manner, so the audio playback on the original device does not stop. 
Moreover, since the {\MOD} issues are only detected though the $\textbf{Test}_\textbf{StartHop}$ test cases, it indicates that the handling process between $\HOP$ operation and $\END$ operation is different. $\END$ operation may force all resources related to the hopping app in the target device to be transferred back to the original device.

\subsubsection{Misuse of Resolution: {\MOR} issues}
As {\MOR} issues involve more statuses, we categorize them into three sub-types according to the status changes, namely {\dtop}, {\dtos}, and {\stopl}.
The first (resp. second) issue refers to the situation where the ``hopping'' resolution for solving audio-stream conflict changes from lowering the volume to playing (resp. stopping) compared to the ``normal'' one. The third issue refers to the situation where the ``hopping''  resolution for solving audio-stream conflict changes from stopping to playing.
Table~\ref{tab:sub-issues}~shows the sub-types of {\MOR} issues {\tool} have detected.
Next, we pick two Navigation apps and a Video app as the case study hopping apps to investigate the characteristic of {\MOR} issue.

\begin{table}[htb]
    \setlength{\abovecaptionskip}{3pt}
    \setlength{\belowcaptionskip}{0pt}
    
    \caption{Sub-types of the App that Detected {\MOR} Issues}\label{tab:sub-issues}
    \begin{center}
    \scalebox{0.85}{ 
    \begin{tabular}{c|c|c|c}
    \hline
        \textbf{App name} & \textbf{\#\dtop}& \textbf{\#\dtos}& \textbf{\#\stopl} \\
    \hline
    QQ Music	& & & $\star$\\
    \hline
    Tencent Video   && &$\star$ \\
    \hline
    AMap        &$\star$  &$\star$ &$\star$\\
    Baidu Map   &$\star$  &$\star$& \\
    Tencent Map &$\star$  &$\star$& \\
    Petal Maps  &$\star$  &$\star$& \\
    \hline
    Xiaohongshu & & & $\star$ \\
    \hline
    \end{tabular}
    }
    \end{center}
    
\end{table}

\textbf{Case study 2:} {\dtop}
type {\MOR}.
When \textit{Baidu Map}~\cite{baidu} is running on the mobile phone and navigating, the user hops it to the tablet for further navigation, on which \textit{Kuaiyin}~\cite{kuaiyin} is playing music.
The expected behaviour is that \textit{Kuaiyin} lower the volume. However, in this situation, both \textit{Baidu Map} and \textit{Kuaiyin} play their audio streams at normal volume on the tablet.
As a result, it makes difficult for users to clearly hear the navigation instructions or information from \textit{Baidu Map}, which brings inconvenience or safety risks to their travels.

\textbf{Case study 3:} {\dtos} type {\MOR}.
When \textit{Petal Map}~\cite{Petal} is running on the mobile phone and navigating, the user hops it to the tablet for further navigation, on which \textit{QQ Music}~\cite{qqmusic} is playing music.
The expected behaviour is that \textit{QQ Music} lower the volume. However, \textit{QQ Music} stops its audio stream. 
On the one hand, it ruins the user's immersive music-listening experience, where the sudden interruption breaks the continuity of the music. On the other hand, the unexpected stop of the music may force the user to interrupt other ongoing operations to check and resume the music playback, distracting the user's attention from using \textit{Petal Map} for navigation or other tasks.

\textbf{Case study 4:} {\stopl} type {\MOR}.
When \textit{Tencent Video}~\cite{Tencent} is playing the video on the mobile phone,
the user hops it to the tablet for further playing, on which \textit{Kugou Music} is playing music.
The expected behaviour is that \textit{Kugou Music} stops playing. However, both \textit{Tencent Video} and \textit{Kugou Music} play their audio streams at normal volume on the tablet.
As a result, users can't clearly distinguish the dialogue in the video from the music, leading to extreme auditory discomfort and ruining the original audio-visual enjoyment.

\textbf{Analysis:} 
After conducting all the experiments, we observed that while \textit{Kuaiyin} and \textit{Kugou Music} exhibit {\MOR} issues as the “pre” apps in hopping, no {\HAC} issues were detected when they served as the “post” apps, i.e., the hopped apps.
Although an \STOP$\rightarrow$\PLAY\  issue was detected in \textit{QQ Music} as shown in Table~\ref{tab:sub-issues}, it occurred in the audio-stream conflict with \textit{Kugou Music}, not with \textit{Tencent Video}. This shows that {\MOR} issues are generally asymmetric, meaning that a change in the order of audio-stream conflict can influence the occurrence of {\MOR} issues.  Thus, we infer that such asymmetric {\MOR} issues may be caused by the fact that apps using different types of audio-streams adopt different priorities for handling audio-stream conflict. 
This indicates that the {\MOR} issues are related to the resolution of conflicts between two apps, which are generally asymmetric. Testers should not design test cases merely based on the conventional symmetric assumption.

\section{Discussion}
This section primarily discusses the threats to the validity (including limited generalizability due to restricted app sampling and version dependency on HarmonyOS) and proposes future research directions (testing in multi-device scenarios, generating test cases with complex hopping operations, and root-cause analysis combining static analysis).
\subsection{Threats to Validity}\label{sec:rq5}
There are two main threats to the validity of our study.
\renewcommand{\labelitemi}{$\bigstar$}
\begin{itemize}[leftmargin=1em]
    \item 
    The representativeness of selected benchmarks can affect the fidelity of our conclusions. To mitigate this threat, we have selected 20 real-world apps from Huawei AppGallery, which are: (1)  Highly popular (Ranked within the top 5 in their respective categories); (2) Diverse in categories (Specifically aligned with audio-stream types, e.g., music players, video platforms, live streaming apps); (3) Large-scale (An avarage of 136.6 MB size). Future work could expand the app sample to include diverse categories and low-download apps, ensuring a more comprehensive assessment of the approach’s robustness.
    \item The version of HarmonyOS, e.g., \textit{3.1}, \textit{4.2}, \textit{Next}, etc., may affect the semantics of app-hopping mechanism. Besides, the latest HarmonyOS version currently faces challenges in experimental validation due to limited device support and a scarcity of available apps, which restricts our ability to assess the framework’s compatibility with emerging system architectures. To mitigate this, we selected the version HarmonyOS 4.2, which is the most widely used version of HarmonyOS up to now, as well as has a large number of available HarmonyOS apps.
\end{itemize}

\subsection{Directions for Further Research}\label{sec:rq4}
\renewcommand{\labelitemi}{$\bigstar$}
According to the previous investigation, we will provide several directions for further researches.
\begin{itemize}[leftmargin=1em]
    \item \textbf{Testing hopping behaviours by generating more complex test cases.} In this paper, the test cases designed are restricted to incorporating only a single hopping operation. However, users may frequently perform multiple consecutive hopping operations. To account for this \seqsplit{real-world} behavior, more complex test cases should be generated in the future, aiming to more comprehensively detect {\HAC} issues.
    \item \textbf{Combining static analysis technique to make in-depth root cause analysis.} In this paper, the cause of {\HAC} issues are analyzed solely based on their phenomena. However, to uncover the root causes of issues, in-depth analysis of the application is required using static analysis techniques to figure out the audio stream related code patterns. Future research works could combine static analysis technique, e.g., data-flow, control-flow, to analyze the root cause of {\HAC} issues.
\end{itemize}

\section{Related work}\label{sec:related}
This section introduces the research works related to HarmonyOS and model-based testing.

\subsection{Analysis and Testing for HarmonyOS}
Since HarmonyOS is an emerging system, there are few research works of analysis and testing for it.
Ma et al.~\cite{MZL23} are the first to provide an overview of HarmonyOS API evolution to measure the scope of situations where compatibility issues might emerge in the HarmonyOS ecosystem.
Zhu et al.~\cite{ZGX23} propose the HM-SAF framework, a cross-layer static analysis framework specifically designed for HarmonyOS applications. The framework analyzes HarmonyOS applications to identify potential malicious behaviors in a stream and context-sensitive manner. 
Chen et al.~\cite{CCY25} design a framework ArkAnalyzer for OpenHarmony Apps.
ArkAnalyzer addresses a number of fundamental static analysis functions that could be reused by developers to implement OpenHarmony app analyzers focusing on statically resolving dedicated issues such as performance bug detection, privacy leaks detection, compatibility issues detection, etc. These works are all static analyses of HarmonyOS apps and do not focus on the {\AC}s studied in this paper.

\subsection{Model-Based Testing of GUI}
Model-based testing (MBT) technique is commonly used in automated GUI testing for applications. Existing woks mainly extract models through static analysis, dynamic analysis and hybrid analysis. FSM~\cite{YPX13} is the first to model the GUI behaviors of Android apps using static analysis for MBT. WTG~\cite{YZW15}, an extension of FSM with back stack and window transition information, is a relatively classic model in MBT. Based on WTG, some models~\cite{GSM19, MHH19, SMC17,CHS18,HCW19,HWC24} which can be considered as a finer-grained WTG, are built by dynamic analysis. There are also some works~\cite{AN13, YWY17,YLP20,LWL22,CLH23} that extend the WTG through a hybrid technique of static and dynamic analysis.
With the rise of large language models (LLMs), the GUI exploration methods based on LLMs are capable of extracting the WTG more quickly and accurately. This new approach leverages the powerful language understanding and generation capabilities of LLMs, which can effectively analyze the complex interactions and transitions within the GUI~\cite{0025CWCWHH024,0025C0CWCW024,abs-2404-02706}.
However, the models proposed in these works are almost used to describe the transitions of GUIs. They do not take into account information related to audio streams, nor do they consider the interactions among multiple applications. These two factors are the key points that {\ASTG} takes into account.

\section{Conclusion}\label{sec:conclusion}
Hopping-related audio-stream conflict (HAC) issues are common on the distributed operating system HarmonyOS. 
To test them automatically and efficiently, we design the Audio Service Transition Graph ({\ASTG}) model and propose a model-based testing approach.
To support it, we also present the first formal semantics of the HarmonyOS's app-hopping mechanism. 
The experimental results show that, with the help of the formal semantics of the app-hopping mechanism and the {\ASTG} model, the {\tool} can detect real-world {\HAC} issues effectively and efficiently.
For the detected issues, we also analyze their characteristics to help app and OS developers improve apps’ quality on distributed mobile systems.

%
 

\section*{Acknowledgement}
This project was partially funded by the Strategic Priority Research Program of the Chinese Academy of Sciences (Grant No. XDA0320102), National Natural Science Foundation of China (Grant No. 62132020), and Major Project of ISCAS (ISCAS-ZD-202302).

\bibliographystyle{ACM-Reference-Format}
\bibliography{main.bib}
\end{document}